# The Boundary Between Privacy and Utility in Data Anonymization

Vibhor Rastogi    Dan Suciu    Sungho Hong

August 19, 2018

## Abstract

We consider the privacy problem in data publishing: given a relation $I$ containing sensitive information "anonymize" it to obtain a view $V$ such that, on one hand attackers cannot learn any sensitive information from $V$, and on the other hand legitimate users can use $V$ to compute useful statistics on $I$. These are conflicting goals. We use a definition of privacy that is derived from existing ones in the literature, which relates the a priori probability of a given tuple $t$, $Pr(t)$, with the a posteriori probability, $Pr(t|V)$, and propose a novel and quite practical definition for utility. Our main result is the following. Denoting $n$ the size of $I$ and $m$ the size of the domain from which $I$ was drawn (i.e. $n < m$) then: when the a priori probability is $Pr(t) = \Omega(n/\sqrt{m})$ for some tuples $t$ there exists no useful anonymization algorithm, while when $Pr(t) = O(n/m)$ for all tuples $t$ then we give a concrete anonymization algorithm that is both private and useful. Our algorithm is quite different from the $k$-anonymization algorithm studied intensively in the literature, and is based on random deletions and insertions to $I$.

## 1 Introduction

The need to preserve private information while publishing data for statistical processing is a widespread problem. By studying medical data, consumer data, or insurance data, analysts can often derive very valuable statistical facts, sometimes benefiting the society at large, but the concerns about individual privacy prevents the dissemination of such databases. Today's state of the art in data anonymization is the $k$-anonymity method [10]: the privacy (or lack thereof) of $k$-anonymity has been studied and improved recently [9, 8, 7, 6, 11], but this method fails to offer any formal guarantees for computing statistical properties, which limits the use of $k$-anonymized data.

Clearly, any anonymization method needs to trade off between privacy and utility: removing all items from the database achieves perfect privacy, but total uselessness, while publishing the entire data unaltered is at the other extreme. In this paper we study the tradeoff between the privacy and the utility of any anonymization method, as a function of the attacker's background knowledge. When the attacker has too much knowledge, we show that no anonymization method can achieve both. But for practical purposes one can often assume that there is a bound on the amount of knowledge the attacker has, and then we show that the tradeoff can be achieved by a new, and very simple anonymization algorithm.

### 1.1 Background

To place our work in context we describe here the problem and some of the issues raised by the $k$-anonymization.

Consider a database $I$ of English test scores as shown in Table 1. It would be useful to make some form of this data publicly available, for example in order to allow researchers in public education to study correlations between various ages, nationalities, and test scores, without releasing the test scores of individual people. While names have already been removed from the table, we don't want to publish this data unchanged because an attacker may use some of the age-nationality values to uniquely identify an individual (this was first shown by Sweeney [10]): for example, he may know that Joe is 21 years old and of Indian Nationality and, since only one entry in the database matches this age and nationality, the attacker can learn Joe's test score. The problem is to *anonymize* the data: more precisely to compute a view $V$ that perturbs the data to make individual identifications impossible. However, we need to do this such as to allow legitimate users to compute answers to "statistical queries". Examples of such queries are: *How many people in the age group 20-30 have score greater*

| Age | Nationality | Score |
|-----|-------------|-------|
| 25  | British     | 99    |
| 27  | British     | 97    |
| 21  | Indian      | 82    |
| 32  | Indian      | 90    |
| 33  | American    | 94    |
| 36  | American    | 94    |

Table 1: A database $I$ of test scores.

*than 90 ?* or *How many people in a particular country have score less than 90 ?* Users evaluate multiple queries like these then combine their answers to derive important correlation between age, score, and nationality: they don't need to see individual scores.

The data anonymization problem has been studied intensively in recent years. Virtually all techniques described in the literature are variations of the $k$-anonymity method [10, 9, 8, 7, 6, 11], which we illustrate next. Note that the anonymization problem is different from the query perturbation problem, studied elsewhere [4, 3, 2].

Figure 2 shows the $k$-anonymity method and several variations described in the literature, for the instance $I$ in Table 1. $k$-anonymization is based on generalizing attribute values, e.g. replacing age 27 with an interval $21 - 30$, or even with a wildcard $*$. Table 2(a) shows the original idea in $k$-anonymization [9]: generalize the attribute values such that every tuple occurs in the data at least $k$ times, in our case $k = 2$. Our hypothetical attacker is no longer able to learn Joe's score, since now there are 2 tuples that match Joe's age and nationality (in general: $k$ tuples). But by generalizing the score values one arguably reduces the utility of the data, and for that reason researchers have proposed to differentiate between the *sensitive attribute* (the score in our example) and the *quasi-identifiers* (age, nationality in our case), and to anonymize only the latter: this is illustrated in our example in Table 2(b), which is still a 2-anonymous table but on the quasi-identifiers only, while the scores are unchanged. One problem with this approach (which is not addressed in the literature) is that it is not always obvious how to classify attributes into quasi-identifiers and sensitive attributes, and this is presumably left to the application: for that reason, in our paper we do not require such a classification of attributes. Continuing our illustration, observe that by keeping the sensitive attribute unaltered one runs the risk of major privacy breaches (as first noted in [8]). This is illustrated in Table 2(b) in the third group: all scores in this group are 94, hence an attacker who knows that (say) Jane is an American can learn that her test score is 94. The solution proposed in [8] is to further require the anonymized data to satisfy a condition called $l$-diversity, namely that in each group there be at least $l$ distinct values of the sensitive attribute. This is shown in Table 2(c), which is 3-anonymous and 2-diverse: within each group, the attacker finds at least 2 (in general $l$) scores that could potentially belong to a particular individual, hence, arguably he cannot guess a private piece of data with probability more than $1/l$.

Thus, most of the prior research on data anonymization has focused on understanding and improving the data's privacy. The formal definitions of privacy in the literature rely on comparing the a priori and the a posteriori probability of a sensitive tuple $t$ being in the database [8]. The a priori probability $Pr[t]$ is the attacker's belief before seeing the published data: for example, the attacker may believe that Joe, who 21 years old and Indian, had a test score of 82 with probability 5%. The a posteriori probability is after seeing the published view, $Pr[t \mid V]$: for example, after seeing the view in Table 2 (b) the a posteriori probability of Joe having a test score of 82 is 50%. When these two probabilities are close, or when the latter is low, then privacy is said to be preserved. In this paper we use a formal definition of privacy that follows the same principles.

Much less understood is the data's utility. Consider the following query: *compute the number of individuals between 26-32 years that received a score $> 90$*: it is unclear how to estimate the query's answer from any of these three tables, and what guarantees this estimate offers. A definition of utility based on entropy is given in [6], but this compares the a priori and the a posteriori distribution, and does not give any guarantees on estimating the answer to a counting query.

| Age | Nationality | Score |
|---|---|---|
| * | British | 96-100 |
| * | British | 96-100 |
| * | Indian | 81-90 |
| * | Indian | 81-90 |
| * | American | 91-95 |
| * | American | 91-95 |

(a) 2-anonymity on every attribute

| Age | Nationality | Score |
|---|---|---|
| * | British | 99 |
| * | British | 97 |
| * | Indian | 82 |
| * | Indian | 90 |
| * | American | 94 |
| * | American | 94 |

(b) 2-anonymity on non-sensitive attributes

| Age | Nationality | Score |
|---|---|---|
| 21-30 | * | 99 |
| 21-30 | * | 97 |
| 21-30 | * | 82 |
| 31-40 | * | 90 |
| 31-40 | * | 94 |
| 31-40 | * | 94 |

(c) 2-diversity and 3-anonymity

| Age | Nationality | Score |
|---|---|---|
| 25 | British | 99 |
| 32 | Indian | 90 |
| 33 | American | 94 |
| 36 | American | 94 |
| 21 | British | 99 |
| 32 | Indian | 89 |
| 28 | Indian | 99 |
| 29 | American | 81 |
| 27 | American | 94 |
| 32 | British | 83 |
| 26 | American | 99 |
| 39 | Indian | 94 |

(d) $(d, \gamma)$-privacy

Table 2: Different methods for computing an anonymized view $V$ for the instance $I$ in Table. 1.

## 1.2 Our Contributions

In this paper we study the privacy/utility tradeoff. For that we give a definition of utility based on estimating *counting queries*, i.e. queries of the form *count the number of records satisfying a certain predicate*. Our definition requires that all counting queries whose answer is "large" be accurately estimated from the anonymized data, with a guarantee on the accuracy that does not depend on any additional assumptions on the original data. For example, if the query *count the number of persons with age between 26 and 32 who received a score over 90* has a large answer (say, 1/50th of the size of the instance), then we require it to be possible to estimate it with high accuracy from the anonymized data. None of the variations of $k$-anonymity described earlier allows us to compute a good estimate to this query: for example in Table 2 (c) all six entries could have an age between 26 and 32, and the only way to estimate the query's answer is to make further assumptions about the distribution of the data: e.g. that ages are uniformly distributed in the interval 21-40, and that the distribution on ages is independent on the test score. But if we ask users to know such properties about the data then we are defeating their very purpose in analyzing the data: the users would like to *discover* whether age and the test score are independent or not, and not to be required to know this in order to discover other facts.

Note that our definition of utility applies only to queries with large answers. This is crucial: queries with small answers could leak privacy if we allow them to be estimated accurately. For example the query *count the number of Indian persons, of age 21, who received a score of 82* has answer 1 on the original data, and if we allowed it to be estimated with high accuracy then privacy would be breached.

Our main result is an almost complete separation of the case when a private/useful anonymization algorithm is possible from that when it is impossible, based on the attacker's knowledge. If the attacker's prior tuple probability is as high as $\Omega(n/\sqrt{m})$, where $n$ is the size of the database and $m$ the size of the domain $D$, then an utility preserving anonymization algorithm is impossible. This impossibility result holds even when the prior distribution is tuple independent. If the attackers attacker's prior is $O(n/m)$, then privacy/utility is possible. We prove that by giving a new, and very simple anonymization algorithm: randomly remove each tuple in $I$ with probability $\alpha$, and randomly insert each tuple in the domain $D$ with probability $\beta$. A particular run of this algorithm is shown in Table 2 (d): Two tuples have been removed from the original data, and eight new tuples have been inserted. (For presentation purposes we indicate which four tuples come from $I$ and which were inserted randomly: in practice this separation is hidden.) Both privacy and utility can be achieved by tuning $\alpha$ and $\beta$, as long as the attacker's prior is $O(n/m)$. Importantly, the accuracy of any counting query is guaranteed as a probability on the random choices made by the algorithm, without any assumptions on the original data. The proof of our algorithm's privacy assumes that the prior distribution is tuple independent, or that it satisfies some very limited form of correlations. This is clearly a weakness, but in some sense it is unavoidable: we show that if the attacker is allowed to know arbitrary tuple correlations, then no utility-preserving algorithm exists even when the prior is $\Omega(n/m)$.

Our impossibility result extends a similar result by Dwork and Nissim [4], which was derived for a different privacy setting. We give the formal definition of the privacy setting in Sec. 3.1. Using this notion of privacy, they showed that no such algorithm can be useful. We improve here, the impossibility result (by tightening the bounds, and relaxing the definition of counting queries), and show how to extend it to the more traditional notion of privacy.

Random deletions and insertions in the database are well known privacy preserving techniques, which have been used in a variety of settings, e.g. in private data mining [5]. Perhaps surprisingly, our algorithm is the first application of these techniques to data publishing.

**Overview** Privacy and utility are defined in Sec. 2; the negative (impossibility) result is in Sec. 3, and the positive result (the new algorithm) is in Sec. 4. Extensions to tuple correlations are discussed in Sec. 5.

## 2 Notations and Definitions

Let $I = \{t_1, t_2, .., t_n\}$ be a database instance where each tuple $t_i$ is drawn from a domain $D$ of size $m$ (i.e. $I \subseteq D$). We need to design a privacy preserving algorithm $A$ which takes as input an instance $I$ and publishes a view $V$, of the same schema as $I$ (i.e. $V \subseteq D$). The view $V$ becomes public knowledge while the original instance $I$ remains hidden.

**Modeling the Adversary** We model the adversary's background knowledge using a probability distribution $Pr_1$ over the set of all possible database instances $I$. Formally, $Pr_1 : 2^D \to [0, 1]$ is such that $\sum_{I \subseteq D} Pr_1[I] = 1$. For each tuple $t \in D$ we denote $Pr_1[t]$ the marginal distribution (i.e. $Pr_1[t] = \sum_{I:t \in I} Pr_1[I]$), and call $Pr_1$ *tuple-independent* when these are independent events. Clearly, we do not know $Pr_1$, only the attacker knows it, and we should design our privacy algorithm assuming the worst about $Pr_1$. As we shall see, however, it is impossible to achieve privacy and utility when the attacker is all powerful, hence we will make reasonable assumptions about $Pr_1$ below.

**Modeling the Algorithm** Given an instance $I$, a privacy-preserving algorithm $A$ will make some random choices and compute a view $V$. We denote $Pr_2^I[V]$ the probability distribution on the algorithm's outputs, i.e. $Pr_2^I : 2^D \to [0, 1]$ is s.t. $\sum_{V \subseteq D} Pr_2^I[V] = 1$. Unlike $Pr_1$, we have total control over $Pr_2$, since we design the algorithm $A$.

As the algorithm $A$ needs to be publicly known, an adversary can compute the induced probability $Pr_{12}[t_i|V]$ based on his prior distribution $Pr_1$ and the algorithm's distribution $Pr_2$, namely $Pr_{12}[t_i|V] = \sum_{I \subseteq D: t \in I} Pr_2^I[V] Pr_1[I]$. We call $Pr_1[t]$ as the prior probability of tuple $t$ while $Pr_{12}[t|V]$ is called the posterior probability of $t$ conditioned on view $V$. Throughout this paper we assume that the adversary is computationally powerful and that $Pr_{12}[t|V]$ can always be computed by the adversary irrespective of the computation effort required.

### 2.1 Classification of Adversaries

Dwork's impossibility result [3] suggest that it is impossible to anonymize data such that it is both useful and protects against arbitrary adversaries. If taken literally, it seems to say that anonymization is hopeless. However, there are lots of examples in practice when data *is* being anonymized and published under risks considered to be acceptable, and this indicates that the privacy definition used by Dwork is too general.

We propose here to study adversaries with restricted power, based on placing a bound on his prior knowledge, and will show that for certain values of this bound privacy is possible.

Specifically, we restrict the adversaries by assuming that for every tuple $t$ his prior probability $Pr_1[t]$ is either small, or equal to 1. In the first case we will seek to hide the tuple from the adversary; in the second case there's nothing to do, since the adversary already knows $t$. Another way to look at this is that we require the algorithm to preserve the privacy for tuples for which $Pr_1[t]$ is small: if $Pr_1[t]$ is large, we may well assume that it is 1, in essence giving up any attempt to hide $t$ from the adversary.

**Definition 2.1.** *Let $d \in (0,1)$. A d-bounded adversary is one for which $\forall t \in D$, either $Pr_1[t] \leq d$ or $Pr_1[t] = 1$.*

We also consider a sub-class of $d$-bounded adversaries called the $d$-independent adversary, which further require $Pr_1$ to be tuple-independent:

**Definition 2.2.** *A d-independent adversary is a d-bounded, tuple-independent adversary.*

## 2.2 Privacy definition and motivation

Our definition of privacy is based on comparing the prior probability $Pr_1[t]$ with the a posteriori probability $Pr_{12}[t \mid V]$, as standard in the literature. We consider here two definitions, one relative and one absolute, and will prove that, in some sense, they are equivalent.

**Definition 2.3.** *An algorithm is called $(d,\delta)$-relative-private if the following holds for all d-independent adversaries $Pr_1$, views $V$, and tuples $t$:*

$$e^{-\delta} \leq \frac{Pr_{12}[t|V]}{Pr_1[t]} \leq e^{\delta}$$

Denoting $de^{\delta} = \gamma$, the definition can be written as:

$$\frac{d}{\gamma} \leq \frac{Pr_{12}[t|V]}{Pr_1[t]} \leq \frac{\gamma}{d}$$

Moreover, if $Pr[t] \leq d$ then $Pr_{12}[t|V] \leq \gamma$, which justifies the following definition of absolute privacy[1]:

**Definition 2.4.** *An algorithm is called $(d,\gamma)$-private if the following holds for all d-independent adversaries $Pr_1$, views $V$, and tuples $t$ s.t. $Pr_1[t] \leq d$:*

$$\frac{d}{\gamma} \leq \frac{Pr_{12}[t|V]}{Pr_1[t]} \qquad Pr_{12}[t|V] \leq \gamma$$

If tuple $t$ fails the left (or right) inequality then we say there has been a negative (or positive) leakage, respectively. Note that positive leakage resembles the notion of positive $(\rho_1, \rho_2)$ breach as described in [5]. We prove the following in the Appendix:

**Proposition 2.5.** *Every $(d,\delta)$-relative-private algorithm is $(d,\gamma)$-private for $\gamma = de^{\delta}$. Conversely, every $(d,\gamma)$-private algorithm is $(d,\delta)$-relative-private for*

$$e^{\delta} = \frac{\gamma}{d}\frac{1-d}{1-\gamma}$$

## 2.3 Utility

As explained earlier, we are interested in supporting counting queries over some predicates. More generally, we define a query $Q$ to be any subset of the domain $D$, i.e. $Q \subseteq D$. The result of the query over the instance $I$ is simply $|Q \cap I|$, which we denote as $Q(I)$. For example, the query *count the number of tuples with age between 26 and 31 and score > 91* is expressed as $Q(I)$, where $Q$ denotes all tuples in the domain having age in $[26, 31]$ score $> 91$.

For any algorithm $A$, using the knowledge of how the algorithm publishes its view one can obtain an estimate for $Q(I)$. Let $EST(Q,V)$ be the estimate[2] of $Q(I)$ as obtained from the published view $V$. Additionally, let $|Q(I) - EST(Q,V)|$ be the absolute error for a query $Q$. We require the algorithm to be such that $EST(Q,V)$ provides a good approximation for $Q(I)$.

---
[1] Only the positive leakage is absolute. For the negative leakage an absolute definition makes no sense.
[2] We assume in this paper that $EST$ is deterministic.

**Definition 2.6.** *A randomized algorithm A is called* useful *if there exists an estimator $EST(Q,V)$ s.t. for any $\epsilon > 0$ there exists a constant $\rho$ such that for all domain size $m$, database size $n$, and query $Q$:*

$$Pr_2 \left[ \, |Q(I) - EST(Q,V)| \geq \rho\sqrt{n} \, \right] \, \leq \, \epsilon$$

## 2.4 Discussion

Our working definition of privacy is $(d, \gamma)$-privacy, where $d$ intuitively corresponds to the prior, and $\gamma$ to the posterior. An intuitive reference point for $d$ is $d = n/m$, since the expected size of $I$ is $n$ for the independent distribution where $Pr_1[t] = n/m$ for all $t$. $\gamma$ is larger, and should be measured in absolute values. For example, $\gamma = 0.05$ means that the view $V$ leaks no tuple $t$ with more than 5% probability.

The motivation for our definition of utility is to give a guarantee on the absolute error of the estimator. It is stated as $\forall \epsilon. \exists \rho$ rather than $\forall \rho. \exists \epsilon$ because the latter is trivially satisfied by any estimator (simply choose $\epsilon = 1$). The absolute error is expressed as $\rho\sqrt{n}$ for two reasons. On one hand we do not want small errors: if one could estimate $Q(I)$ with an error $< 1$ then there would be privacy breaches, as we have seen. On the other we do not want big errors: *any* estimator is accurate with probability 1 if the absolute error is $n$. In between 1 and $n$ we have chosen $\rho\sqrt{n}$.

$k$-anonymity and $l$-diversity satisfy neither $(d, \gamma)$-privacy nor usefulness. Considering a $d$-independent adversary, privacy is compromised when the adversary knows that some tuples have very *low* probability. To see the intuition, consider the instance in Table 2(c). Suppose a $d$-independent adversary is trying to find out Joe's test score, and he knows that Joe is likely to have a low score: i.e. the prior is such that $Pr_1[t]$ is very low for a tuple saying that Joe's score is greater than 95, and larger (but still $\leq d$) for tuples saying that his score less than 90. If the adversary knows that Joe's age is less than 30, then his record is among the first three: since the first two tuples have a very *low* probability (as their scores are 99, 97), the adversary concludes that Joe's scores is 82 with very *high* probability. There is no utility either. Suppose we want to estimate the number of students between 29 and 31 years old from Table 2(c): the answer can be anywhere between 0 and 6, and, if our estimate is the average, 3, then the only way we can guarantee any accuracy is by making assumptions on the distribution of the data, in essence by knowing $Pr_1$.

By contrast, the algorithm described in Sec. 4 takes two constants $k, \gamma$ and is both $(k\frac{n}{m}, \gamma)$-private and also useful: for any $\epsilon$, it gives $\rho = 4\sqrt{3\frac{k}{\gamma}ln(\frac{2}{\epsilon})}$.

## 3 Impossibility Results

We prove here that no $(d, \gamma)$-private algorithm can provide even a weak notion of utility if $d = \Omega(n/\sqrt{m})$. For that we first establish an impossibility result for a weaker notion of privacy, called $\epsilon$-indistinguishability, and a very weak notion of utility, called meaningfulness: in this form, our result is an improvement of the impossibility result in [4]. Then we establish an impossibility result for our notions of privacy and utility.

### 3.1 The Strong Impossibility Result

Consider the following alternative definition of privacy [4]:

**Definition 3.1.** *An algorithm is $\epsilon$-indistinguishable if for all database instances $I$ and $I'$ which disagree exactly over a pair of tuples (i.e. $|I| = |I'|$ and $|I - I'| = 2$) and for all views $V$,*

$$e^{-\epsilon} \leq \frac{Pr_2^I[V]}{Pr_2^{I'}[V]} \, \leq \, e^{\epsilon}$$

The weak notion of utility, which we call meaningfulness was first considered in [4] and is based on the notion of statistical difference:

**Definition 3.2.** *The statistical difference between two distributions $Pr_A$ and $Pr_B$ over the domain $X$ is $SD(Pr_A, Pr_B) = \sum_{x \in X} | Pr_A(x) - Pr_B(x) |$.*

Note that $SD(Pr_A, Pr_B) \in [0,2]$: it is 0 when $Pr_A = Pr_B$, and is 2 when $\forall x.(Pr_A(x) = 0 \lor Pr_B(x) = 0)$. We explain now informally the connection between statistical difference and utility (the formal connection is in Prop 3.6 below). Suppose an algorithm $A$ gives reasonable estimates to counting queries $Q$. Let $Q$ be a "large" query, i.e. if executed on the entire domain it returns sizeable fraction, say $1/5$. When we estimate $Q$ on the view published for a particular instance $I$ we will get some errors, which depend on actual size $n = |I|$. If there is any utility to $A$ then a user should be able distinguish between the two extreme cases, when $Q(I) = n$ and when $Q(I) = 0$. To capture this intuition, define the uniform distribution $Pr_1$: $\forall I$ s.t $|I| = n$, $Pr_1[I] = \frac{1}{\binom{m}{n}}$. Note that $\binom{m}{n}$ is the number of instances of size $n$. Thus $Pr_1$ makes every instance of size $n$ equally likely. Let $E_Q$ be the event $(|I| = n \land Q(I) = n)$, and $E'_Q$ the event $(|I| = n \land Q(I) = 0)$. Then we expect to be able to differentiate between the following two distributions: $Pr_A^Q = Pr_{12}[V|E_Q]$ and $Pr_B^Q = Pr_{12}[V|E'_Q]$. To obtain a reasonable estimate for $Q$, $SD(Pr_A^Q, Pr_B^Q)$ should be large. On the other hand, if $SD(Pr_A^Q, Pr_B^Q)$ is very small then no reasonable estimate of the query can be obtained from any of the published views. An algorithm is meaningless if the $SD(Pr_A^Q, Pr_B^Q)$ is small for a large fraction of the queries $Q$. An algorithm is meaningful if it is not meaningless.

**Definition 3.3.** *Let $f < 1$ be a constant independent of the domain size $m$ and database size $n$. Consider all queries $Q$ s.t $\frac{1}{2}(1-f) \leq \frac{|Q|}{m} \leq \frac{1}{2}(1+f)$. An algorithm $A$ is called meaningless if $SD(Pr_A^Q, Pr_B^Q)$ is smaller than $1/2$ for a fraction $2/3$ of queries $Q$.*

Next we state the strong impossibility result using meaningfulness as the definition of utility. For the sake of concreteness we use the constant $1/2$ for statistical difference and $2/3$ for the fraction of queries in the definition of meaningfulness. However, the impossibility result works for arbitrary constants.

**Theorem 3.4** (Strong Impossibility). *There exists a constant $c$ independent of $m$ and $n$ such that every algorithm which satisfies $\epsilon$-indistinguishability with $e^\epsilon \leq c\frac{m}{n^2}$ is meaningless.*

Before giving the proof, we comment on how this result extends a similar result in [4]. First, we have generalized it to a larger class of queries: the previous result restricts the class of selection queries to certain `xor` operations, here we allow arbitrary selection queries. Secondly, we improve the bound on the statistical difference. This was possible because the original proof relies on a chaining argument which provides a bound on the statistical difference of a function at each step of the chain to eventually compute the statistical difference of $Pr_A$ and $Pr_B$. At each step it considers tuples as points in a high dimensional space and bounds the statistical difference of a function which satisfies $\epsilon$-indistinguishability over tuples. We observe that each database instance can be thought of as a point in a higher dimensional space and thus bound the statistical difference of a function which satisfies $\epsilon$-indistinguishability over instances.

**Proof of Theorem 3.4:** Consider all instances $I$ such that $|I| = n$. The number of such instances with all distinct tuples is $\binom{m}{n}$. Moreover, as $A$ satisfies $\epsilon$-indistinguishability, for any pair of instances $I$ and $I'$ where $|I| = |I'| = n$ and $|I - I'| = 2$,

$$e^{-\epsilon} \leq \frac{Pr_2^I[V]}{Pr_2^{I'}[V]} \leq e^\epsilon$$

We denote by $Pr_{12}[V]$ the probability that the view $V$ is published. If $Pr_1$ is the uniform distribution then

$$Pr_{12}[V] = \frac{1}{\binom{m}{n}} \sum_{|I|=n} Pr_2[V|I]$$

Consider a query $Q$ s.t $|Q| = m\tau$ where $\frac{1}{2}(1-f) \leq \tau \leq \frac{1}{2}(1+f)$. Let us represent the set of instances for which $Q(I) = n$ as $S_Q$. Then $|S_Q|$ is $\binom{m\tau}{n}$. Let $E_Q$ be the event that input database instance $I$ belongs to $S_Q$. $Pr_{12}[V|E_Q]$ is the probability that the view $V$ is published if an instance $I$ is picked uniformly at random from the set $S_Q$.

$$Pr_{12}[V|E_Q] = \frac{1}{|S_Q|} \sum_{I \in S_Q} Pr_2[V|I]$$

If we consider all queries $Q$ s.t $|Q| = m\tau$, then we see that the expectation $\mathbf{E}_Q(Pr_{12}[V|E_Q]) = Pr_{12}[V]$. We show in Lemma B.1 that if $e^\epsilon \leq \frac{m\tau - 2n}{2n^2}$ then the variance of $Pr[V|E_Q]$ over $Q$ of fixed size $m\tau$ is small and is less than $\frac{8e^\epsilon Pr_{12}[V]^2 n^2}{m\tau}$. Using this we can show that the statistical difference between the distributions $Pr_{12}[V]$ and $Pr_{12}[V|E_Q]$ is small with high probability over choice of $Q$. As shown in Lemma B.3, with probability at least $1 - \alpha$,

$$SD(Pr_{12}[V], Pr_{12}[V|E_Q]) = O\left(\frac{e^\epsilon n^2}{\alpha m\tau}\right)^{\frac{1}{3}}$$

Fix $\alpha = \frac{1}{6}$. As $\tau \geq \frac{1}{2}(1+f)$ and $f$ is a constant independent of $n$ and $m$, with probability greater than $\frac{5}{6}$, $SD(Pr_{12}[V], Pr_{12}[V|E_Q])$ is $O\left(\frac{e^\epsilon n^2}{m}\right)^{\frac{1}{3}}$. The same holds for the statistical difference between $Pr_{12}[V]$ and $Pr_{12}[V|E'_Q]$. Thus with probability greater than $\frac{2}{3}$ over the choice of $Q$,

$$SD(Pr_{12}[V|E_Q], Pr_{12}[V|E'_Q]) = O\left(\frac{e^\epsilon n^2}{m}\right)^{\frac{1}{3}}$$

Thus there exists a constant $c$ such that if $e^\epsilon \leq c\frac{m}{n^2}$ then $SD(Pr_{12}[V|E_Q], Pr_{12}[V|E'_Q]) \leq \frac{1}{2}$ for at least 2/3 of the queries $Q$ making the algorithm meaningless. This completes the proof.

## 3.2 Impossibility for $(d, \gamma)$-Privacy

We show now how the strong impossibility result translates to our notions of privacy/utility. First, we show that $(d, \delta)$-relative-privacy implies $\epsilon$-indistinguishability. Note that the former is a privacy notion about an adversary, while the latter does not talk about any adversary. The proof of the following theorem is in the appendix.

**Proposition 3.5.** *Every $(d, \delta)$-relative-private algorithm satisfies $\epsilon$-indistinguishability with $\epsilon = 2\delta + 2\ln(2)$*

Next, we connect the two notions of utility:

**Proposition 3.6.** *Any useful algorithm is also meaningful.*

**Proof:** Consider the distributions $Pr_A^Q, Pr_B^Q$ defined above, for the events $E_Q = (Q(I) = n) \wedge (|I| = n)$ and $E'_Q = (Q(I) = 0) \wedge (|I'| = n)$. Since the algorithm is useful, we will choose the value $\epsilon = \frac{1}{4}$ in Definition 2.6, and obtain a value for $\rho$. Let $I_0$ be the set of instances $I$ such that $Q(I) = 0$ and $|I| = n$. Similarly, let $I_n$ be the set of instances $I'$ such that $Q(I') = n$ and $|I'| = n$. Let $V_0$ be the set of views $V$ such that $EST(Q, V) \leq \rho\sqrt{n}$. Similarly, let $V_n$ be the set of views $V'$ such that $EST(Q, V') \geq n - \rho\sqrt{n}$.

As $A$ is useful, a view $V \in V_0$ would be published on any $I \in I_0$ with probability greater than $\frac{3}{4}$. Thus, $Pr_{12}[V \in V_0|E_Q] \geq \frac{3}{4}$. Similarly, $Pr_{12}[V \in V_n|E'_Q] \geq \frac{3}{4}$

If $n \geq 2\rho\sqrt{n}$ then $V_0$ and $V_n$ have to be disjoint. As $\rho$ is independent of $n$, we can choose a large enough $n$ such that $\sqrt{n} \geq \rho$. For those values of $n$, the statistical difference between $Pr_A^Q$ and $Pr_B^Q$ is at least $2(\frac{3}{4} - \frac{1}{4}) = 1$, for every $Q$. Thus any useful algorithm is also meaningful

**Corollary 3.7.** *Let $A$ be a meaningful algorithm and, let $\gamma < 1$ be any given number. Then there exists a constant $c$ independent of $m$, $n$ and $\gamma$ such that there is a d-independent adversary with $d = \frac{1}{c}\frac{\gamma}{1-\gamma}\frac{n}{\sqrt{m}}$ for which one of the following is true for some tuple $t$*

- *There is positive leakage: $Pr_1[t] \leq d$ but $Pr_{12}[t|V] \geq \gamma$*

- *There is negative leakage: $\frac{Pr_{12}[t|V]}{Pr_1[t]} \leq \frac{1}{c}\frac{n}{\sqrt{m}}$*

**Proof:** We can define for any algorithm

$$e^\epsilon = \max_{I,I'} \frac{Pr_2[V|I]}{Pr_2[V|I']}$$

for all $I$ and $I'$ which disagree in a single pair of tuples. By Theorem 3.4, the algorithm can have meaningful utility only if $e^\epsilon = \Omega(\frac{m}{n^2})$. Using Proposition 3.5, we can infer that if the algorithm is meaningful, it cannot satisfy $(d, \delta)$-relative-privacy unless $e^\delta \geq c\frac{\sqrt{m}}{n}$, for some constant $c$ independent of $n$ and $m$. Thus, no meaningful algorithm can satisfy $(d, \gamma)$-privacy unless $\frac{\gamma}{d}\frac{1-d}{1-\gamma} \geq c\frac{\sqrt{m}}{n}$. As $A$ is meaningful, at least one of the following statements is true:

- There is positive leakage: $Pr_1[t_i] \leq d$ but $Pr_{12}[t_i|V] \geq \frac{1}{1+\frac{1}{cd}\frac{n}{\sqrt{m}}}$.
  In particular if $d = \frac{1}{c}\frac{\gamma}{1-\gamma}\frac{n}{\sqrt{m}}$ then $Pr_{12}[t_i|V] \geq \gamma$

- There is negative leakage: $\frac{Pr_{12}[t_i|V]}{Pr_1[t_i]} \leq \frac{d}{\gamma} \leq \frac{d}{\gamma}\frac{1-\gamma}{1-d} \leq \frac{1}{c}\frac{n}{\sqrt{m}}$

This completes the proof.

## 4 Algorithm

As we have seen, it is impossible to guarantee absence of privacy breaches against all $d$-independent adversaries if $d = \Omega(\frac{n}{\sqrt{m}})$. In this section, we present a simple algorithm which is $(d, \gamma)$-private. Here, $d$ and $\gamma$ are parameters given to the algorithm. The utility of the algorithm depends on the values of $d$ and $\gamma$ and a guarantee for the utility can be made only for $d = k\frac{n}{m}$, for any fixed constant $k$ (hence $d = O(\frac{n}{m})$).

Assume that the input database instance is $I \subseteq D$. The insert-remove algorithm computes the view $V$ as following:

- For every tuple in $I$, insert it in the $V$ independently with probability $\alpha$
- For every tuple in $D$ which is not in $I$, insert it in the view $V$ independently with probability $\beta$
- Publish $D$, $V$, $\alpha$, $\beta$.

### 4.1 Privacy Analysis

**Theorem 4.1.** *The insert-remove algorithm is $(d, \gamma)$-private where $d \leq \gamma$ if we choose $\alpha \leq 1 - \frac{d}{\gamma}$ and $\beta \geq \frac{d}{\gamma}(\frac{1-\gamma}{1-d})\alpha$*

**Proof:** Consider any tuple $t$. Let $Pr[t] = p \leq d$. We know that

$$Pr[t|V] = \frac{Pr[V|t]Pr[t]}{Pr[V|t]Pr[t] + Pr[V|\bar{t}]Pr[\bar{t}]}$$
$$= \frac{Pr[V|t]p}{Pr[V|t]p + Pr[V|\bar{t}](1-p)}$$

If $d \leq \gamma$ then $\beta \leq \alpha$. Now consider the following two cases:

- $t$ is present in the published view. Thus $Pr[V|t] = \alpha$ and $Pr[V|\bar{t}] = \beta$. Using this we get,

$$Pr[t|V] = \frac{\alpha p}{\alpha p + \beta(1-p)}$$
$$\leq \frac{\alpha d}{\alpha d + \alpha d\frac{1-\gamma}{\gamma}} \leq \gamma$$

On the other hand, for negative leakage, note that $\frac{Pr[t|V]}{Pr[t]} \geq \frac{\alpha}{\beta} \geq 1 \geq \frac{d}{\gamma}$.

- $t$ is not present in the published view. Thus $Pr[V|t] = 1-\alpha$ and $Pr[V|\bar{t_i}] = 1-\beta$. Using this we get,

$$Pr[t|V] = \frac{(1-\alpha)p}{(1-\alpha)p + (1-\beta)(1-p)}$$

As $\alpha \geq \beta$, we get $\frac{\alpha}{\beta} \geq \frac{1-\alpha}{1-\beta}$ and

$$\frac{(1-\alpha)p}{(1-\alpha)p + (1-\beta)(1-p)} \leq \frac{\alpha p}{\alpha p + \beta(1-p)}$$

As $\frac{\alpha p}{\alpha p + \beta(1-p)}$ is an increasing function of $\frac{\alpha}{\beta}$ for a fixed p. Thus even in this case we have $Pr[t|V] \leq \gamma$. Additionally, $\frac{Pr[t|V]}{Pr[t]} \geq \frac{(1-\alpha)}{(1-\beta)} \geq 1 - \alpha \geq \frac{d}{\gamma}$.

## 4.2 Utility Analysis

The estimator $EST(Q,V)$ is the following. Recall that $\alpha$ and $\beta$ are published.

- Let $n_V = Q(V)$; that is $n_V = |Q \cap V|$ is the query evaluated on the view $V$.
- Let $n_D = Q(D)$; that is $n_D = |Q|$ is the query evaluated on the entire domain (we explain below how to do this efficiently)
- Define:

$$EST(Q,V) = \frac{n_V - \beta n_D}{\alpha - \beta}$$

**Theorem 4.2.** *Let r be a constant and $\alpha, \beta$ be such that $\alpha \geq \frac{1}{2}$ and $\beta \leq \frac{r}{4}\frac{n}{m}$. Then for any $\epsilon > 0$, denoting $\rho = 2\sqrt{3rln(\frac{2}{\epsilon})}$ we have:*

$$Pr_2\left[\ |Q(I) - EST(Q,V)| \geq \rho\sqrt{n}\ \right] \leq \epsilon$$

*It follows that the algorithm is useful for $\alpha \geq \frac{1}{2}$ and $\beta \leq \frac{r}{4}\frac{n}{m}$.*

**Proof:** Let $\mu_V$ be the expected number of tuples in $V$ which satisfy the query $Q$. Then,

$$\mu_V = \alpha Q(I) + \beta(n_D - Q(I))$$

which reduces to

$$Q(I) = \frac{\mu_V - \beta n_D}{\alpha - \beta}$$

Instead of using $\mu_V$, we use $n_V$ and obtain $EST(Q,V)$. Thus, the absolute error $\Sigma = |Q(I) - EST(Q,V)|$ is

$$\Sigma = \frac{|\mu_V - n_V|}{\alpha - \beta}$$

Using Chernoff bound we can say that with an probability $1 - 2e^{-\frac{\mu_V \delta^2}{3}}$, the fractional error $\frac{|\mu_V - n_V|}{\mu_V} \leq \delta$. Thus with probability $\geq 1 - 2e^{-\frac{\mu_V \delta^2}{3}}$, we know that $\Sigma \leq \frac{\delta \mu_V}{\alpha - \beta}$. Hence, for $\delta = \sqrt{\frac{3ln\frac{2}{\epsilon}}{\mu_V}}$, with probability greater than $1 - \epsilon$ the following holds:

$$\Sigma \leq \sqrt{3ln\frac{2}{\epsilon}}\frac{\sqrt{\mu_V}}{\alpha - \beta}$$

$$\leq \sqrt{3ln\frac{2}{\epsilon}}\sqrt{\frac{Q(I)}{\alpha} + \frac{\beta n_D}{\alpha^2}}$$

As $\alpha = \frac{1}{2}$ and $\beta \leq \frac{r}{4}\frac{n}{m}$, we can guarantee that the error $\Sigma$ would be less than $2\sqrt{3rln(\frac{2}{\epsilon})n}$ with probability greater than $1 - \epsilon$. Since $\epsilon$ was arbitrary, the algorithm is useful. This completes the proof.

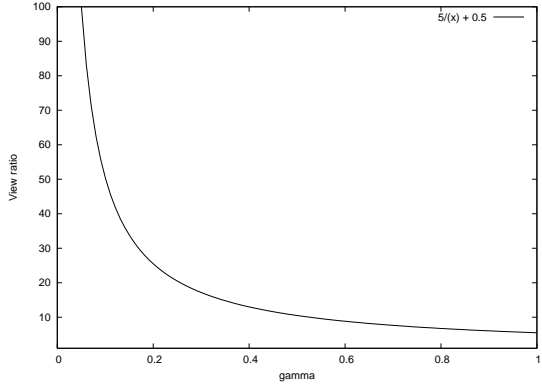

Figure 1: View Size Ratio vs $\gamma$

Assuming that $d \leq \frac{\gamma}{2}$, for satisfying the trade off between privacy and utility we can choose $\alpha = \frac{1}{2}$ and a suitable $\beta$, if

$$\frac{d}{\gamma}(\frac{1-\gamma}{1-d})\alpha \leq \frac{r}{4}(\frac{n}{m})$$

This is true when $\frac{d}{\gamma} \leq \frac{r}{4}\frac{n}{m}$. Thus for $d = k\frac{n}{m}$, the algorithm satisfies $(d,\gamma)$-privacy and is also useful, as can be seen by setting $\rho = 4\sqrt{3\frac{k}{\gamma}ln(\frac{2}{\epsilon})}$ for $\epsilon > 0$.

## 4.3 Discussion

**Choice of parameters** Suppose we want to ensure $(k\frac{n}{m}, \gamma)$-privacy, where $k$ is a constant, i.e. $d = k\frac{n}{m}$ is the bound on the adversary's prior probability, and $\gamma$ is the bound on the posteriori probability that is acceptable for us. We also want to ensure utility. Choosing $\alpha = \frac{1}{2}$ satisfies both Theorems 4.1 (assuming $d/\gamma < 1/2$) and 4.2. Next, we choose $\beta = \frac{k}{\gamma}\frac{n}{m} = \frac{d}{\gamma}$. This satisfies Theorem 4.1 (since $1 > \frac{1-\gamma}{1-d}\alpha$), and also satisfies Theorem 4.2 whenever $\frac{k}{\gamma} \leq \frac{r}{4}$. In other words, with these choices of parameters we have $(d,\gamma)$-privacy, and a utility that is captured by $\rho = 2\sqrt{3rln(\frac{2}{\epsilon})}$, where $r = \frac{4k}{\gamma}$ (smaller $r$'s are better for utility). The privacy/utility tradeoff is like this. We want to protect against a powerful adversary, i.e. large $k$: we want to protect well, i.e. small $\gamma$; and this limits the utility expressed by $r$.

**The View Size** A concern about this algorithm is the potential size increase of the view: since most of the privacy comes from the newly inserted tuples (i.e. $\beta$), clearly $V$ will be larger than $I$. The expected size of $V$ is simply $n\alpha + (m-n)\beta$. For the chosen values of $\alpha = \frac{1}{2}$ and $\beta = \frac{k}{\gamma}\frac{n}{m}$ the formula reduces to $n(\frac{1}{2} + \frac{k}{\gamma})$ which means a similar tradeoff exists for the view size. Fig 1 shows the tradeoff by plotting the relationship between size ratio of the view and $\gamma$ for a fixed $k = 5$. The size ratio of a view $V$ is simply $\frac{|V|}{|I|}$.

**The Domain** If $I$ is a table with $a$ attributes, then we chose as domain $D$ the cross product of all active domains of all attributes of $I$, i.e. $D = D_1 \times \ldots \times D_a$, where $D_i = \pi_i(I)$, $i = 1, \ldots, a$. This has two implications. First, the algorithm computes each $D_i$, and publishes $D_i$ separately; each $m_i = |D_i|$ is relatively small, while the size of the entire domain $m = m_1 m_2 \ldots m_a$ is huge. The large size of the domain means that the estimator needs to compute $Q(D) = |Q \cap D|$ efficiently. This is easy, since $Q$ is usually a conjunction of predicates over several attributes, hence can be computed on each $D_i$ independently. Secondly, in the insertion phase of the algorithm tuples from the domain are added by first computing the number of tuples to be added using a binomial distribution with parameters $m - n$ and $\beta$. Then a sample of that size is drawn by randomly selecting tuples from the domain without replacement.

**A Case Study** To understand the tradeoff between privacy and utility, we experimented with US census data in the Adults database which has also been used for experiments in ([8],[1]). The database $I$ has 9 attributes and $n = 30162$ tuples. The product of the attributes active domains has $m = 648023040$ tuples, thus $n/m \approx 4.65 \times 10^{-3}$. We chose the parameters $\alpha$ and $\beta$ to ensure $(10n/m, 0.2)$-privacy: i.e. the adversary's prior is bounded by $d = 0.465\%$ and the posterior by $\gamma = 20\%$. The upper bound on the posteriori, $\gamma$, as a theoretical upper bound, is for an arbitrarily biased adversary: in practice, the posteriori is much smaller. We chose the algorithm's parameters as $\alpha = 0.5$ and $\beta = 9.5 \times 10^{-4}$.

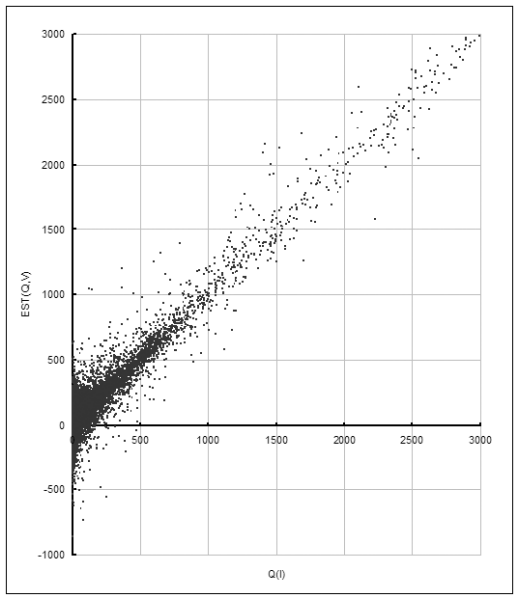

Figure 2: EST(Q,V) vs Q(I)

Next we tested the utility of this data, by running all possible selection queries with up to three attributes. Figure 2 shows the result. Each dot represents one query $Q$, where $x = Q(I)$ and $y = EST(V, Q)$: thus, a perfect estimator would correspond to the $y = x$ line. This experiments illustrates the desired behavior of our algorithm. For small values of $Q(I)$ the estimated values are far off (even negative): this is necessary to preserve privacy. But for large values of $Q(I)$ the estimations are much better and meaningful. As expected most of the values lie in a band of width 1000 around the line $y = x$, thus showing that the error was indeed additive in nature.

## 5 Tuple correlations

So far our analysis was restricted to tuple-independent adversaries. This restriction strengthens the impossibility result, and weakens that of the algorithm. Here we discuss several extensions to tuple correlations. First, we show that if the adversary knows arbitrary correlations, then no algorithm can achieve both privacy and utility. Then we examine a restricted class of correlations, which is sufficiently general to model join/link attacks: we show that our algorithm still guarantees privacy against *positive* leakages (utility, of course, is unchanged), but it cannot protect against negative leakage. The policy of protecting against positive leakage while permitting negative leakage is sometimes acceptable in practice, and it raises the question whether our algorithm could be strengthened in this case: after all the result in Corollary 3.7 relies on both negative and positive leakages. We answer this negatively, by giving a variant of the impossibility result for positive leakages only.

### 5.1 Arbitrary Correlations

Suppose we publish a view $V$ of a database of diseases with schema (`age`, `zip`, `disease`) using our algorithm. An attacker knows the age and zipcode for both Joe and Jim. Now Joe and Jim are brothers: if one has diabetes, then the other is quite likely to have diabetes as well (at least that's what the attacker believes). Suppose the attacker finds two tuples in $V$ matching both Joe and Jim having diabetes. The probability that none was in $I$ and were inserted by the algorithm is very small: $\beta^2$. In contrast, because of their strong correlation, the probability that they were in the instance $I$ is now much larger: $Pr_1[t \cap t'] \gg Pr_1[t]Pr_1[t']$. Thus, upon seeing both tuples the attacker concludes that they are Joe and Jim's with high probability. This is the reason why our algorithm has difficulties hiding data when the prior has correlations.

We show here that even for $d \geq \frac{n}{m}$ no private and useful algorithm exists if the adversary is allowed to know arbitrary correlations. The proof appears in the appendix.

**Theorem 5.1.** *If $A$ is a useful algorithm, then for any $\epsilon > 0$ there exists $n$ such that for every $d \geq \frac{n}{m}$ there is a $d$-bounded adversary and tuple $t$ such that $Pr_1[t] \leq d$ but $Pr_{12}[t|V] \geq 1 - 2\epsilon$*

## 5.2 Exclusions

We consider now some restricted forms of correlations: exactly one tuple from a set of tuples occurs in the database.

**Definition 5.2.** *A $d$-exclusive adversary is a $d$-bounded adversary with the following kind of correlations among the tuples: There is a partition of $D$ into a family of disjoint sets, $D = \bigcup_j S_j$, s.t.*

- *Tuples are pairwise independent if they do not belong to the same set.*
- *Exactly one tuple in each set occurs in the database instance*

This type of correlations model naturally adversaries performing join/link attacks. Such attackers are able to determine some of the identifying attributes of a particular tuple, say the age and nationality of a person that they know must occur in the database. The adversary can thus identify a set of tuples $S_j$ of the domain such that exactly one tuple in the set belongs to the database.

### 5.2.1 Positive results for $d$-exclusive adversaries

We have already seen the privacy analysis for the insert-remove algorithm against $d$-independent adversaries. In this section, we show that the algorithm provides a slightly weaker form of guarantee for $d$-exclusive adversaries. We show that the algorithm ensures that there is no positive leakage for negatively correlated $d$-bounded adversaries: That is if $Pr_1[t] \leq d$ then $Pr_{12}[t|V] \leq \gamma$.

**Theorem 5.3.** *If $\alpha = \frac{1}{2}$ and $\beta \geq \frac{d}{\gamma}(\frac{1-\gamma}{1-d})$ then the algorithm ensures absence of positive leakage for all tuples and for all $d$-exclusive adversaries, i.e if $Pr_1[t] \leq d$ then $Pr_{12}[t|V] \leq \gamma$*

The proof of the theorem appears in the appendix. The algorithm cannot guarantee absence of negative leakage for $d$-exclusive adversaries, as seen from the following example:

**Example 5.4** Suppose we publish a view $V$ of a database of diseases with schema (`age,zip,disease`) using our algorithm. An attacker knows the age and zipcode of a person Joe and additionally knows that no other person in the database has the same combination of age and zipcode. The prior belief of the attacker is that Joe has exactly one disease but any one of the diseases is equally likely. Suppose the attacker finds a tuple $t_1$ in $V$ matching Joe and Diabetes. Let us consider the tuple $t_2$ corresponding to Joe and Malaria. Theorem 5.3 shows that there won't be a positive leakage for $t_1$. However, for $t_2$ there is a drastic drop in the a posteriori probability causing a negative leakage.

### 5.2.2 Impossibility results for $d$-exclusive adversaries

In this section we extend the impossibility result for $d$-exclusive adversaries to accommodate the case when negative leakage is legally allowed and only positive leakage is considered as a privacy breach. The impossibility result shows that if an algorithm satisfies some form of weak utility then there exists a $d$-exclusive adversary, for $d = \Omega(\frac{n}{\sqrt{m}})$, which can infer that a certain tuple exists in the database with high probability.

The extension works only for a restricted class of randomized algorithms. However, the class is broad enough to encompass many of the privacy preserving algorithms in the literature. The class of algorithms which we consider satisfy the following bucketization assumptions:

- The algorithm is such that if the prior distribution is tuple independent then the posterior distribution is also tuple independent. More formally, the tuples in the domain can be partitioned into buckets such that the distribution $Pr_{12}[t_i|V]$ over the tuples obeys the following condition: If two tuples $t_i$ and $t_j$ lie in different buckets then they are independent conditioned on the view.

- Let the number of buckets be $N_B$. We assume that the distribution of tuples both of the database instance and the domain among the buckets is not too skewed. More formally, for every $k > 1$ there exists a $N_k < N_B$ such that if we remove any $N_k$ buckets, the remaining still contain a fraction $1/k$ of tuples of $I$ as well as $D$.

For example consider the method of full domain generalization (in [7]) as applied to ensure $k$-anonymity or the anatomy method (in [11]) as applied to ensure $l$-diversity. For both the methods the tuples corresponding to the different anonymized groups form buckets which satisfy the assumptions above.

To show the impossibility result for $d$-independent adversaries, we used $\epsilon$-indistinguishability as the privacy definition and meaningfulness as the utility definition. Impossibility result for $d$-exclusive adversaries requires a slight modification to both privacy and utility definitions.

**Definition 5.5.** *Consider all database instances $I$ and $I'$ which disagree exactly over a pair of tuples (i.e. $|I| = |I'|$ and $|I - I'| = 2$). An algorithm satisfies $\epsilon$-indistinguishability over a set $D'$, if for all database instances $I$ and $I'$ which disagree exactly over a pair of tuples with both the tuples in $D'$ and for all views $V$,*

$$e^{-\epsilon} \leq \frac{Pr_2^I[V]}{Pr_2^{I'}[V]} \leq e^{\epsilon}$$

Thus $\epsilon$-indistinguishability over a set $D'$ is a relaxation of the original $\epsilon$-indistinguishability definition. Next we consider the notion of utility called $k$-meaninglessness. The notion of utility is a generalization of meaninglessness and is also defined for counting queries $Q$ s.t $\frac{1}{2}(1-f) \leq \frac{|Q|}{m} \leq \frac{1}{2}(1+f)$, for a constant $f < 1$. Intuitively, the definition tries to capture the fact that an algorithm will have bad utility if many queries $Q$ have an error of $O(n)$ in their estimates.

**Definition 5.6.** *An algorithm is called $k$-meaningless, if there is a set $S$ with $|S| \leq n(1-\frac{1}{k})$ such that the distributions $Pr_A'^Q$ and $Pr_B'^Q$ have a statistical difference smaller than $1/2$ for a fraction $2/3$ of the queries $Q$.*
$Pr_A'^Q$: $Pr_{12}[V|E_Q]$ where $E_Q$ is the event $Q(I) = |Q \cap S|$ and $|I| = n$
$Pr_B'^Q$: $Pr_{12}[V|E_Q']$ where $E_Q'$ is the event $Q(I) = |Q \cap S| + \frac{n}{k}$ and $|I| = n$

Note that for $k = 1$, the definition reduces to notion of meaninglessness. $k$-meaninglessness follows the same intuition: If $SD(Pr_A'^Q, Pr_B'^Q)$ is small then it would be impossible to distinguish whether the original answer is $|Q \cap S|$ or $|Q \cap S| + \frac{n}{k}$ thus resulting in an error of at least $\frac{n}{k}$. Cor. 5.7 relates the notions of $\epsilon$-indistinguishability over the set $D'$ and $k$-meaninglessness; the proof appears in the appendix.

**Corollary 5.7.** *Let $D'$ be a set of tuples such that $|D'| \geq \frac{m}{k}$ and $|D' \cap I| \geq \frac{n}{k}$, for some constant $k$ independent of $m$ and $n$. Then there exists a constant $c$ such that every algorithm which satisfies $\epsilon$-indistinguishability over the set $D'$ with $e^\epsilon \leq c\frac{m}{n^2}$ is $k$-meaningless.*

We call any algorithm which is not $k$-meaningless as $k$-meaningful. As any algorithm which is meaningless is also $k$-meaningless, it implies that every $k$-meaningful algorithm is meaningful.

**Theorem 5.8.** *Let $A$ be $k$-meaningful algorithm which satisfies the bucketization assumptions with $N_k \geq \frac{3}{d}$. Then there exists a constant $c$ independent of $n$ and $m$ for which there is a $d$-exclusive adversaries with $d = max(\frac{3}{N_k}, c\frac{n}{\sqrt{m}})$ having positive leakage on some tuple.*

The proof of Theorem 5.8 appears in the appendix. We give a brief overview outlining the intuition: We argue that for a $k$-meaningful algorithm which satisfies the bucketization assumptions, there exists a $d'$-independent adversary for which either there is a positive leakage on one tuple or there is negative leakage on lot of tuples. The presence of multiple negative leakages is shown by using the fact that, for a $k$-meaningful algorithm, any large set containing a significant fraction of tuples of the database will have leakage on some tuple. Using the bucketization assumption we prove that if there is no positive leakage, then many buckets will contain tuples having negative leakage. In case of positive leakage we are done as the $d'$-independent adversary serves as $d$-exclusive adversaries. In the case of multiple negative leakages, we explicitly construct a $d$-exclusive adversary using the $d'$-independent adversary such that there is positive leakage for at least one tuple. Here, we use $d' = d/3$.

# 6 Conclusions

We have described a formal framework for studying both the privacy and the utility of an anonymization algorithm. We proved a tight bound between privacy and utility, based on the attacker's power. For the case where privacy/utility can be guaranteed, we have described a new, quite simple anonymization algorithm, based on random insertions and deletions of tuples in/from the database. We have done a limited empirical study, and saw a good privacy/utility tradeoff. Our algorithm increases the size of the data, but by tolerable amounts (a factor of 10, in our empirical study). It will be interesting to study in future work ways to reduce the size of the published view.

# A Relationships among privacy definitions

## A.1 $(d, \gamma)$-privacy and $(d, \delta)$-relative-privacy

**Proposition** (2.5). *Every $(d, \delta)$-relative-private algorithm is $(d, \gamma)$-private for $\gamma = de^\delta$. Conversely, every $(d, \gamma)$-private algorithm is $(d, \delta)$-relative-private for*

$$e^\delta = \frac{\gamma}{d}\frac{1-d}{1-\gamma}$$

**Proof:** The proof of the first part trivially follows from the definitions. For the converse let $A$ be a $(d, \gamma)$-private algorithm. Thus $A$ ensures no positive/negative leakage for all $d$-independent adversaries. We shall use this fact to show that $A$ should satisfy $(d, \delta)$-relative-privacy. For this we consider $Pr_{12}[t_i|V]$ purely as a function of $Pr_1[t_1]$ while keeping the prior probability of all other tuples constant. Let us call the function as $f$. We compute the function explicitly as:

$$Pr_{12}[t_i|V] = \frac{Pr_{12}[V \cap t_i]}{Pr_{12}[V \cap t_i] + Pr_{12}[V \cap \bar{t_i}]} = \frac{\Sigma_{I_j \in I_t} Pr_{12}[V \cap I_j]}{\Sigma_{I_j \in I_t} Pr_{12}[V \cap I_j] + \Sigma_{I'_j \in I_{\bar{t}}} Pr_{12}[V \cap I'_j]]}$$

$$= \frac{\Sigma_{I_j \in I_t} Pr_2[V|I_j] Pr_1[I_j]}{\Sigma_{I_j \in I_t} Pr_2[V|I_j] Pr_1[I_j] + \Sigma_{I'_j \in I_{\bar{t}}} Pr_2[V|I'_j] Pr_1[I'_j]}$$

Here, $I_t$ represent the set of instances which contain $t_i$ and $I_{\bar{t}}$ represent the set which does not. For each instance $I_j$ in the set $I_t$ we can decompose $Pr_1[I_j]$ as $Pr_1[t_i]Pr_1[T_j]$ where $T_j$ is the event that denotes that only the tuples in $I_j$ except for $t_i$ occur in the database. This decomposition is possible because of tuple independence in $Pr_1$. Using this decomposition we can rewrite $Pr_{12}[t_1|V]$ as

$$\frac{c_1 Pr_1[t_i]}{c_1 Pr_1[t_i] + c_2(1 - Pr_1[t_i])}$$

where $c_1$ and $c_2$ are constants as we vary $Pr_1[t_i]$ and keep the prior probabilities of all other tuples constant. Let us represent $Pr_{12}[t_i|V]$ as the function

$$f(x) = \frac{c_1 x}{c_1 x + c_2(1-x)}.$$

We notice that the function is increasing in $x$ and has slope $\frac{c_1}{c_2}$ at the origin. Additionally, if $c_1 \geq c_2$ then $f(x) \geq x$ for all $x$. On the other hand if $c1 < c_2$ then $f(x) \leq x$ for all $x$. We are interested in the maximum and minimum possible values of $\frac{f(x)}{x}$. The maximum occurs when $c_1 > c_2$ at points near the origin and is equal to $\frac{c_1}{c_2}$. The fact that $A$ has be safe from privacy leakages imposes certain restrictions on $f$. One condition is that $f(d) \leq \gamma$ which implies that $\frac{c_1}{c_2} \leq \frac{\gamma(1-d)}{(1-\gamma)d}$. Another is that $\forall x : \frac{f(x)}{x} \geq \frac{d}{\gamma}$. It follows that such an algorithm $A$ is also $(d, \delta)$-relative-private for $e^\delta = \max(\frac{\gamma(1-d)}{(1-\gamma)d}, \frac{\gamma}{d}) = \frac{\gamma(1-d)}{(1-\gamma)d}$.

## A.2 $(d, \delta)$-relative-privacy and $\epsilon$-indistinguishability

For proving Proposition 3.5, we need to show the following lemma,

**Lemma A.1.** *Let $A$ be a $(d, \delta)$-relative-private algorithm. Then, for all tuples $t_i$ such that $Pr_1[t_i] \leq d$,*

$$\frac{1 - de^\delta}{1-d} \leq \frac{Pr_{12}[\bar{t_i}|V]}{Pr_1[\bar{t_i}]} \leq \frac{1 - \frac{d}{e^\delta}}{1-d} \tag{1}$$

**Proof:** Assume $Pr_1(t_i) \leq d$. Additionally, we know that

$$e^{-\delta} \leq \frac{Pr_{12}[t_i|V]}{Pr_1[t_i]} \leq e^\delta$$

It is easy to see that using the above two inequalities we can bound $\frac{Pr_{12}[\bar{t}_i|V]}{Pr_1[\bar{t}_i]}$ as required. This completes the proof.

Using sandwich theorem on (1), we note that

$$lim_{d\to 0}(\frac{Pr_{12}[\bar{t}_i|V]}{Pr_1[\bar{t}_i]}) = 1$$

Thus, for any given $\delta$, we can choose $d$ small enough such that $\frac{1}{2} \leq \frac{Pr_{12}[\bar{t}_i|V]}{Pr_1[\bar{t}_i]} \leq 2$

**Proposition** (3.5). *Every $(d,\delta)$-relative-private algorithm satisfies $\epsilon$-indistinguishability with $\epsilon = 2\delta + 2\ln(2)$*

**Proof:** Consider a $d'$ for which $Pr'_1$ is such that

$$\frac{1}{2} \leq \frac{Pr'_{12}[\bar{t}_i|V]}{Pr'_1[\bar{t}_i]} \leq 2 \qquad (2)$$

We know that such a $d'$ exists for every $\delta$. If $d \leq d'$ then we let $Pr'_1 = Pr_1$. On the other hand if $d > d'$, then we use the fact any $(d,\delta)$-relative-private algorithm would also be $(\delta, d')$-private algorithm and hence from here on we consider the algorithm as $(\delta, d')$-private and assume that equation (2) holds true for $Pr_1$.

Consider any two database instances $I$ and $I''$ which differ in exactly one tuple with $I$ containing one extra tuple $t_1$. Then we can represent $I$ as $t_1 \cap T$ and $I''$ as $\bar{t}_1 \cap T$. Here $T$ represents the event that all tuples in $I''$ belong to the database and all tuples in $\bar{I}$ do not belong to the database.

$$\frac{Pr_2[V|I]}{Pr_2[V|I'']} = \frac{Pr_2[V|(t_1 \cap T)]}{Pr_2[V|(\bar{t}_1 \cap T)]} = \frac{Pr_{12}[V \cap t_1 \cap T]}{Pr_{12}[V \cap \bar{t}_1 \cap T]}\frac{Pr_1[\bar{t}_1]}{Pr_1[t_1]} = \frac{Pr_{12}[t_1|(V \cap T)]}{Pr_{12}[\bar{t}_1|(V \cap T)]}\frac{Pr_1[\bar{t}_1]}{Pr_1[t_1]}$$
$$= \frac{Pr_{12}[t_1|(V \cap T)]}{Pr_{12}[t_1|V]}\frac{Pr_{12}[\bar{t}_1|(V \cap T)]}{Pr_2[\bar{t}_1|V]}\frac{Pr_{12}[t_1|V]}{Pr_1[t_1]}\frac{Pr_{12}[\bar{t}_1|V]}{Pr_1[\bar{t}_1]}$$

Consider a $d$-independent adversary for which $Pr_1[T] = 1$ but $Pr_1[t_1] \leq d$. As the algorithm satisfies $(d,\delta)$-relative-privacy, it should be safe against such an adversary. Thus,

$$e^{-\delta}(\frac{1}{2}) \leq \frac{Pr_2[V|I]}{Pr_2[V|I'']} \leq e^{\delta}(2)$$

Similarly consider the database instances $I'$ and $I''$ such that $I'$ has one extra tuple $t_2$. Again, we observe that

$$e^{-\delta}(\frac{1}{2}) \leq \frac{Pr_2[V|I']}{Pr_2[V|\bar{I''}]} \leq e^{\delta}(2)$$

Thus, combining the two inequalities we can see that there exists an $\epsilon = 2(\delta + ln(2))$ such that

$$e^{-\epsilon} \leq \frac{Pr_2[V|I']}{Pr_2[V|I]} \leq e^{\epsilon}$$

This completes the proof.

# B   The strong Impossibility result

**Lemma B.1.** *if $e^{\epsilon} \leq \frac{m\tau - 2n}{2n^2}$ then $\mathbf{Var}_Q(Pr_{12}[V|E_Q]) \leq \frac{8e^{\epsilon}Pr_{12}[V]^2 n^2}{m\tau}$*

**Proof:** Let $x$ denote a database instance of size $n$ and let $p(x)$ denote $Pr_{12}[V|x]$. Thus, $Pr_{12}[V]$ is exactly $\mathbf{E}_x p(x) = \sum_x \frac{p(x)}{\binom{m}{n}}$. Additionally, $Pr_{12}[V|E_Q] = \mathbf{E}_{x \in S_Q} p(x) = \sum_{x \in S_Q} \frac{p(x)}{\binom{m\tau}{n}}$. Note that $\mathbf{E}_Q(Pr_{12}[V|E_Q]) = Pr_{12}[V]$. We want to compute $\mathbf{Var}_Q(Pr_{12}[V|E_Q])$. Let $\chi_Q(x)$ denote the indicator variable corresponding to the event $x \in S_Q$.

$$
\begin{aligned}
\mathbf{Var}_Q(Pr_{12}[V|E_Q]) &= \mathbf{E}_Q \left( \frac{\sum_x p(x)\chi_Q(x)}{\binom{m\tau}{n}} \right)^2 - \left( \frac{\sum_x p(x)}{\binom{m}{n}} \right)^2 \\
&= \frac{1}{\binom{m\tau}{n}} \sum_{i=0}^n \sum_{|x-y|=2i} \frac{p(x)p(y)}{\binom{m}{n}\binom{m-n}{i}\binom{n}{n-i}} \left[ \binom{m\tau-n}{i}\binom{n}{n-i} - \binom{m-n}{i}\binom{n}{n-i}\frac{\binom{m\tau}{n}}{\binom{m}{n}} \right] \\
&= \frac{1}{\binom{m\tau}{n}} \sum_{i=0}^n a_i(c_i - d_i) \\
&= \frac{1}{\binom{m\tau}{n}} \sum_{i=0}^n a_i b_i
\end{aligned}
$$

Here $|x-y|$ is the symmetrical difference between the instances $x$ and $y$. Thus $|x-y|$ represents the number of tuples in which $x$ and $y$ disagree. We use the notation $a_i = \sum_{|x-y|=2i} \frac{p(x)p(y)}{\binom{m}{n}\binom{m-n}{i}\binom{n}{n-i}}$, $c_i = \binom{m\tau-n}{i}\binom{n}{n-i}$, $d_i = \binom{m-n}{i}\binom{n}{n-i}\frac{\binom{m\tau}{n}}{\binom{m}{n}}$ and $b_i = c_i - d_i$.

Note that $\sum_i b_i = 0$. Moreover we can rewrite $b_i$ as the product of $\binom{m-n}{i}\binom{n}{n-i}$ and $\frac{\binom{m\tau-n}{i}}{\binom{m-n}{i}} - \frac{\binom{m\tau}{n}}{\binom{m}{n}}$. We can see that the first factor is always positive while the second factor decreases as $i$ increases. Thus $b_i \leq 0 \implies b_{i+1} \leq 0$.

For the sequence $a_i$ it is easy to see that $\sum_i a_i \frac{\binom{m-n}{i}\binom{n}{n-i}}{\binom{m}{n}} = \sum_x \frac{p(x)^2}{\binom{m}{n}^2} = Pr_{12}[V]^2$. Due to property of $\epsilon$-indistinguishability we know that if $|x-y|=1$ then $e^{-\epsilon} \leq \frac{p(x)}{p(y)} \leq e^\epsilon$. Using this we show in Lemma B.2 that $\forall i, e^{-\epsilon} \leq \frac{a_i}{a_{i+1}} \leq e^\epsilon$.

Let $i_0$ be the largest index s.t. $b_{i_0} > 0$. Then $b_{i_0+1} \leq 0$ and thus $\frac{\binom{m\tau-n}{i_0+1}}{\binom{m\tau}{n}} \leq \frac{\binom{m-n}{i_0+1}}{\binom{m}{n}}$. As $\sum_{i=0}^n a_i \frac{\binom{m-n}{i}\binom{n}{n-i}}{\binom{m}{n}} = Pr_{12}[V]^2$, we know that in particular

$$a_{i_0+1} \frac{\binom{m-n}{i_0+1}\binom{n}{n-i_0-1}}{\binom{m}{n}} \leq Pr_{12}[V]^2$$

$$\implies \frac{a_{i_0}}{e^\epsilon} \frac{\binom{m-n}{i_0+1}\binom{n}{n-i_0-1}}{\binom{m}{n}} \leq Pr_{12}[V]^2$$

$$\implies \frac{a_{i_0}}{e^\epsilon} \frac{\binom{m\tau-n}{i_0+1}\binom{n}{n-i_0-1}}{\binom{m/2}{n}} \leq Pr_{12}[V]^2$$

$$\implies a_{i_0} \frac{\binom{m\tau-n}{i_0}\binom{n}{n-i_0}}{\binom{m\tau}{n}} \frac{(m\tau-n-i_0)(n-i_0)}{(i_0+1)^2} \leq e^\epsilon Pr_{12}[V]^2$$

$$\implies a_{i_0} \frac{\binom{m\tau-n}{i_0}\binom{n}{n-i_0}}{\binom{m\tau}{n}} \leq \frac{n^2 e^\epsilon Pr_{12}[V]^2}{m\tau-2n}$$

$$\implies \frac{a_{i_0} c_{i_0}}{\binom{m\tau}{n}} \leq \frac{2n^2 e^\epsilon Pr_{12}[V]^2}{m\tau-2n}$$

If $e^\epsilon \leq \frac{m\tau-2n}{2n^2}$ then $a_i c_i \leq \frac{a_{i+1}c_{i+1}}{2}$ for all $i$. Thus the entire sum $\frac{1}{\binom{m\tau}{n}} \sum_{i=0}^n a_i b_i$ can be bounded as $\frac{1}{\binom{m\tau}{n}} \sum_{i=0}^{i_0} a_i c_i \leq \frac{2 a_{i_0} c_{i_0}}{\binom{m\tau}{n}} \leq \frac{4n^2 e^\epsilon Pr_{12}[V]^2}{m\tau-2n}$. As $m$ is much larger than $n$ and $\tau$ is a constant, we can assume that $m\tau \geq 4n$. Thus we get that $\mathbf{Var}_Q(Pr_{12}[V|E_Q]) \leq \frac{8n^2 e^\epsilon Pr_{12}[V]^2}{m\tau}$. This completes the proof.

**Lemma B.2.** $\forall i, e^{-\epsilon} \leq \frac{a_i}{a_{i+1}} \leq e^{\epsilon}$.

**Proof:** Let $S_i^x$ be the set of instances $\{y \mid |x-y| = 2i\}$

$$
\begin{aligned}
a_i &= \sum_{|x-y|=2i} \frac{p(x)p(y)}{\binom{m}{n}\binom{m-n}{i}\binom{n}{n-i}} \\
&= \sum_x \frac{p(x)}{\binom{m}{n}} \left[ \sum_{y \in S_i^x} \frac{p(y)}{\binom{m-n}{i}\binom{n}{n-i}} \right]
\end{aligned}
$$

For each $y$, consider the set of instances $S_y$ obtained by removing one of the $i$ tuples in $y$ by one of the $i$ tuples of $x$. Hence there are $i^2$ such instances. Note that $\forall y' \in S_y, p(y') \leq e^{\epsilon} p(y)$. On the other hand each instance $y' \in \cup_{y \in S_i^x} S_y$ is obtained from exactly $(n-i+1)(m-n-i+1)$ instances from $S_i^x$. Thus

$$
\begin{aligned}
a_i &\geq \sum_x \frac{p(x)}{\binom{m}{n}} \left[ \sum_{y' \in S_{i-i}^x} \frac{(n-i+1)(m-n-i+1)}{i^2 \binom{m-n}{i}\binom{n}{n-i}} \frac{p(y')}{e^{\epsilon}} \right] \\
&\geq \sum_x \frac{p(x)}{\binom{m}{n}} \left[ \sum_{y' \in S_{i-i}^x} \frac{p(y')}{e^{\epsilon}\binom{m-n}{i-1}\binom{n}{n-i+1}} \right] = \frac{a_{i-1}}{e^{\epsilon}}.
\end{aligned}
$$

Similarly we can show that $a_i \leq e^{\epsilon} a_{i-1}$. This completes the proof

**Lemma B.3.** *With probability at least* $(1-\alpha)$, $SD(Pr[z], Pr_S[z]) \leq O\left(\frac{e^{\epsilon} n^2}{\alpha m \tau}\right)^{\frac{1}{3}}$

**Proof:** The proof is exactly the same as in the proof of Lemma 2 shown in [4].

# C Tuple correlations

## C.1 Arbitrary correlations

**Theorem.** *5.1 If A is a useful algorithm, then for any $\epsilon > 0$ there exists $n$ such that for every $d \geq \frac{n}{m}$ there is a $d$-bounded adversary and tuple $t$ such that $Pr_1[t] \leq d$ but $Pr_{12}[t|V] \geq 1 - 2\epsilon$*

**Proof:** Let $I$ be the input instance and $n = |I|$. Consider a $d$-bounded adversary with the following tuple correlations: The adversary knows that either all tuples of the set $S = I$ belong to the database or none of the tuples belong to the database. Consider the query $Q = S$. As the algorithm is useful, for every $\epsilon$ there exists a $\rho$ such that the $|EST(Q,V) - Q(I)| \leq \rho\sqrt{n}$ with probability greater than $\epsilon$. If $n$ is large enough, then it follows that algorithm cannot output the same view on $I$ and any instance $I'$ in the set $\{I'|S \subset D - I'\}$ with probability greater than $2\epsilon$. Hence, from the view there will be a breach with probability greater than $1 - 2\epsilon$.

## C.2 $d$-exclusive adversaries

### C.2.1 Positive result for $d$-exclusive adversaries

**Theorem** (5.3). *If $\alpha = \frac{1}{2}$ and $\beta \geq 2\frac{d}{\gamma}(\frac{1-\gamma}{1-d})$ then the insert-remove algorithm ensures absence of positive leakage for all tuples and for all $d$-exclusive adversaries, i.e if $Pr_1[t] \leq d$ then $Pr_{12}[t|V] \leq \gamma$*

**Proof:** The probability $Pr[t_i|V]$ is only dependent on which tuples in $S$ appear in $V$. It is clear that maximum privacy leakage happens when $V$ contains $t_i$ but does not contain any other tuple from $S$. In that case, $Pr[V|t_i] =$ $\alpha(1-\beta)^{(|S|-1)}$ and $Pr[V|t_j] = \beta(1-\alpha)(1-\beta)^{(|S|-2)}$. Also, let $Pr[t_i] = p$. For a $d$-exclusive adversary $\bar{t}_i = \cup_{j \neq i} t_j$. Thus we get,

$$Pr[t_i|V] = \frac{\alpha(1-\beta)p}{\alpha(1-\beta)p + \beta(1-\alpha)(1-p)}$$

As $p \leq d$, $\alpha = \frac{1}{2}$ and $\beta \geq \frac{d}{\gamma}(\frac{1-\gamma}{1-d})$, we get $Pr[t_i|V] \leq \gamma$. This completes the proof.

## C.3  Impossibility result for $d$-exclusive adversaries

**Corollary** (5.7). *Let $D'$ be a set of tuples such that $|D'| \geq \frac{m}{k}$ and $|D' \cap I| \geq \frac{n}{k}$, for some constant $k$ independent of $m$ and $n$. Then there exists a constant $c$ such that every algorithm which satisfies $\epsilon$-indistinguishability over the set $D'$ with $e^\epsilon \leq c\frac{m}{n^2}$ is $k$-meaningless.*

**Proof:** We can think of a random $Q$ with $|Q| = \frac{m(1-f)}{2}$ as picking a random subset of size $\frac{m(1-f)}{2}$ from $D$. Let $X$ stand for the random variable denoting the number of tuples in $D'$ which are also in $Q$. Then $\mathbf{E}(X) = \frac{|D'|(1-f)}{2}$ and $\mathbf{Var}(X) = \frac{|D'|(1-f)}{4}$. If $|D'| > \frac{m}{k}$ then using Chernoff inequality we can show that with probability greater $1 - 2e^{-\frac{m}{16k}}$ the random subset will have size between $\frac{m(1-f)}{2k}$ and $\frac{3m(1-f)}{2k}$ in $D'$.

Let $I' = D' \cap I$, then using Theorem 3.4 for tuples in $I'$ we can show that with probability greater than $\frac{2}{3}$, the statistical difference between Distribution 0' and Distribution 1' is at most $O\left(\frac{e^\epsilon n^2}{m}\right)^{\frac{1}{3}} + 2e^{-\frac{m}{16k_2}}$ which is dominated by $O\left(\frac{e^\epsilon n^2}{m}\right)^{\frac{1}{3}}$.

Distribution 0': $Pr_{12}[V|E]$ where $E$ is the event $Q(I') = 0$
Distribution 1': $Pr_{12}[V|E']$ where $E'$ is the event $Q(I') = |I'|$

This shows the distributions $Pr'^Q_A$ and $Pr'^Q_B$ have statistical difference $O\left(\frac{e^\epsilon n^2}{m}\right)^{\frac{1}{3}}$ with probability greater than $\frac{2}{3}$, where
$Pr'^Q_A$: $Pr_{12}[V|E_Q]$ where $E_Q$ is the event $Q(I) = |Q \cap I'|$
$Pr'^Q_B$: $Pr_{12}[V|E'_Q]$ where $E'_Q$ is the event $Q(I) = |Q \cap I'| + \frac{n}{k}$
If $e^\epsilon = O(\frac{m}{n^2})$, then $A$ is $k$-meaningless.

**Theorem** (5.8). *Let $A$ be $k$-meaningful algorithm which satisfies the bucketization assumptions with $N_k \geq \frac{2}{d}$. Then there exists a constant $c$ independent of $n$ and $m$ for which there is a $d$-exclusive adversary with $d = \max(\frac{3}{N_k}, c\frac{n}{\sqrt{m}})$ having positive leakage on some tuple.*

**Proof:** As $A$ is $k$-meaningful, it is also meaningful. By Corollary 3.7 we know that there exists a $d'$-independent adversary $Ad_1$ which either has a positive leakage or negative leakage. Here we use $d'$ such that $d' = \frac{d}{3}$. If it has positive leakage then that adversary serves as the required $d$-exclusive adversary. On the other hand, if it has negative leakage on tuple $t_1$ we choose the bucket $B_1$ which contains that tuple. The set $D' = D - B_1$ is thus the set of tuples which are conditionally independent of tuple $t_1$.

Let us define, $I' = I \cap D'$. We know from the bucketization assumptions that $|D'| \geq \frac{|D|}{k}$ and $I' \geq \frac{|I|}{k}$. Let us define, $e^\epsilon = \max_{I,I'} \frac{Pr_2[V|I]}{Pr_2[V|I']}$ for all $I$ and $I'$ which disagree over a single pair of tuples with both of them in $D'$. As $A$ is $k$-meaningful, from Corollary 5.7, we know that $e^\epsilon = \Omega(\frac{m}{n^2})$. By restricting over the set $D'$ and using Corollary 3.7, it follows that there exists a $d'$-independent adversary $Ad'_2$ for which there is a leakage for some tuple $t_2$ in the set $D'$. If there is positive leakage for $t_2$ then $Ad'_2$ serves as the required $d$-exclusive adversary. If there is negative leakage on $t_2$ then let the bucket which contains it be $B_2$. Consider the adversary $Ad_2$ which has prior of $Ad'_2$ for tuples in $B_2$ and prior of $Ad_1$ for tuples in $B_1$. As the posterior probabilities of tuples in $B_2$ are independent of that of tuples in $B_1$ even conditioned on $V$, the leakages for $Ad'_2$ in $B_2$ and for $Ad_1$ in $B_1$ would be preserved for $Ad_2$. Thus, $Ad_2$ has negative leakage for both tuples $t_1$ and $t_2$.

We repeat this procedure $\frac{3}{d}$ times, with buckets being $B_i$ for $i \in [1, \frac{3}{d}]$. Define $B = \cup_i B_i$ and $D_B = D - B$. As $N_k \geq \frac{3}{d}$, by the bucketization assumptions on the algorithm, it follows $|D_B| \geq \frac{m}{k}$ and $|D_B \cap I| \geq \frac{n}{k}$. Hence,

Corollary 5.7 holds for each step and we can show that there is either a $d'$-independent adversary for which there is positive leakage on some tuple or there exists a $d'$-independent adversary $Ad$ for which there is negative leakage on at least $\frac{3}{d}$ tuples. In the former case, we have the required $d$-exclusive $d$-bounded adversary. For the latter case, let the set of tuples with negative leakage be $S'$. We know that each tuple in $S'$ belongs to a different bucket. Thus, we can increase the prior probability of each tuple $t_i$ in $S$ so that $Pr_1[t_i] = d'$. The negative leakage on each $t_i \in S$ is still preserved because as shown in the proof of Proposition 2.5, the ratio $\frac{Pr_{12}[t_i|V]}{Pr_1[t_i]}$ increases by at most $\frac{1}{1-d'}$ if we increase $Pr_1[t_i]$ to $d'$ while keeping the prior probability of all other tuples constant. For $d' \leq \frac{1}{2}$, the increase in the ratio is by a factor of at most 2, thus preserving the negative leakage.

Additionally there exists at least one tuple for which the prior $Pr_1$ of $Ad$ is such that $Pr_{12}[t_p|V] = cd'$. This is because $D_B$ contains at least $\frac{m}{k}$ tuples and $Ad$ can have any prior probabilities for each one of them as they will never effect the posterior probabilities of the leaking tuples. So, we can change the prior of any tuple in $D'$ so as to get the required posterior probability for $t_p$. This is shown in Lemma C.1 proved in the appendix. Let us define $S$ as $S' \cup t_p$

Using $Ad$ we construct a $d$-exclusive adversary $Ad'$ having positive leakage on tuple $t_p$. $Ad'$ corresponds to the adversary who knows exactly one tuple in $S$ occurs in the database instance. We construct $Ad'$ from $Ad$ by making the prior probability of each illegal instance (instances which do not exactly one tuple from $S$) as 0. We do this by distributing the probability of each such illegal instance over the entire set of legal instances such that the probability of every legal instance increases by an amount proportional to its original probability. Let $L$ denote the set of legal instances and $IL$ denote set of illegal instances.

For $Ad$, Let us call $\tau = \sum_{I' \in IL} Pr_1^{Ad}[I']$. We know that $\tau = 1 - (1-d')^{(\frac{1}{d'}-1)} \leq 1 - \frac{1}{e}$. In $Ad'$ this sum gets distributed to every legal instance and thus $Pr_1^{Ad'}[I] = Pr_1^{Ad}[I](\frac{1}{1-\tau})$ for every legal instance I. We know that $Pr_1^{Ad'}[t_i] \leq Pr_1^{Ad}[t_i]\frac{1}{1-\tau} \leq d'e \leq d$. Thus $Ad'$ is $d$-bounded. Lemma C.2, proved in the appendix, shows that the constructed adversary $Ad'$ is indeed $d$-exclusive. We next show that there is a positive leakage on tuple $t_p$ for $Ad'$. Let us compute for any $t_i$ in $S'$ and $t_p$ the ratio

$$\frac{Pr_{12}^{Ad'}[t_i|V]}{Pr_{12}^{Ad'}[t_p|V]} = \frac{\sum_{I \supset t_i} Pr_2[V|I]Pr_1^{Ad'}[I]}{\sum_{I' \supset t_i} Pr_2[V|I']Pr_1^{Ad'}[I']} \quad (3)$$

$$= \frac{\sum_{I \supset t_i} Pr_2[V|I]Pr_1^{Ad}[I](\frac{1}{1-\tau})}{\sum_{I' \supset t_i} Pr_2[V|I']Pr_1^{Ad}[I'](\frac{1}{1-\tau})} \quad (4)$$

$$= \frac{\sum_{I \supset t_i} Pr_2[V|I]Pr_1^{Ad}[I]}{\sum_{I' \supset t_i} Pr_2[V|I']Pr_1^{Ad}[I']} = \frac{Pr_{12}^{Ad}[t_i|V]}{Pr_{12}^{Ad}[t_p|V]} \quad (5)$$

$$\leq \frac{\frac{n}{\sqrt{m}}d'}{cd} \leq \frac{1}{c'}\frac{n}{\sqrt{m}} \quad (6)$$

Note that (4) can be derived from (3) as both the summation are being done over a subset of legal instances. Also (6) is true as there is negative leakage for $t_i$ and $Pr_{12}^{Ad}[t_p|V] = cd$. Note that (6) holds for all $t_i$ in $S'$. Moreover, we know that

$$\sum_{i \in S'} Pr_{12}^{Ad'}[t_i|V] + Pr_{12}^{Ad'}[t_p|V] = 1$$

Thus $Pr_{12}^{Ad'}[t_p|V]$ is at least $\frac{1}{1+\frac{1}{c'd}\frac{n}{\sqrt{m}}}$. Hence there exists a $d$-exclusive adversary for which there is positive leakage for at least one tuple. This completes the proof.

**Lemma C.1.** *Let A be a k-meaningful algorithm. Let $D' \subset D$ be a set of tuples such that $|D'| \geq \frac{m}{k}$ and $|D' \cap I| \geq \frac{n}{k}$. Then there exists a tuple t in $D'$ such that $Pr_{12}^u[t|V] \geq c'\frac{n}{m}$ for the uniform tuple independent distribution $Pr_1^u$ and some constant $c'$ independent of n and m.*

**Proof:** We know that for any query $Q$,

$$EST(Q,V)_D = \sum_{t_i \in D} Pr_{12}^u[t_i|V]$$

If we restrict the EST(Q,V) over tuples in $D'$, then

$$EST(Q,V)_{D'} = \sum_{t_i \in D'} Pr_{12}^u[t_i|V]$$

Let us define $I' = D' \cap I$. For the algorithm to have some utility over domain $D'$, there exists a constant $l$ with $l > k$ for which some query $Q$ has the following properties: $|Q \cap I'| \geq \frac{n}{k}$ and $EST(Q,V)_{D'} \geq \frac{n}{l}$. If not then it would be impossible to distinguish whether $Q(I) \geq \frac{n}{k}$ or $Q(I) \leq \frac{n}{l}$ from the view for every query $Q$. As $EST(Q,V)_{D'} = \sum_{t_i \in D'} Pr_{12}^u[t_i|V] \geq \frac{n}{l}$. This means, there exists a tuple $t$ in $D'$ such that $Pr_{12}^u[t|V] \geq \frac{n}{lm}$. Hence proved.

Note that $\frac{Pr_{12}^u[t|V]}{Pr_1^u[t]} = \frac{1}{l}$. As shown in the proof of Proposition 2.5 increasing the prior probability to $d$ will change the ratio of posterior to prior probability to $\frac{d}{d+l} \leq \frac{d}{2l}$. Thus, for the tuple $t$ there exist $Pr_1$ such that $Pr_{12}[t|V] \geq cd$.

**Lemma C.2.** *The constructed adversary $Ad'$ is d-exclusive*

**Proof:** We show that $Ad'$ is $d$-exclusive adversary by first showing that all tuples $t_1$ and $t_2$ which are not in $S$ are still independent. Let us define $E$ as the event that exactly one tuple in $S$ occurs in the database. We know that

$$\begin{aligned} Pr_1^{Ad}[t_1 \cap t_2 \cap E] &= \frac{Pr_1^{Ad}[t_1 \cap E] Pr_1^{Ad}[t_2 \cap E]}{Pr_1^{Ad}[E]} = \frac{(\sum_{I \in LI | I \supset t_1} Pr_1^{Ad}[I])(\sum_{I \in LI | I \supset t_2} Pr_1^{Ad}[I])}{\sum_{I \in IL} Pr_1^{Ad}[I]} \\ &= (1-\tau) \frac{(\sum_{I \in LI | I \supset t_1} Pr_1^{Ad'}[I])(\sum_{I \in LI | I \supset t_2} Pr_1^{Ad'}[I])}{\sum_{I \in IL} Pr_1^{Ad'}[I]} \\ &= (1-\tau) \frac{Pr_1^{Ad'}[t_1 \cap E] Pr_1^{Ad'}[t_2 \cap E]}{Pr_1^{Ad'}[E]} \end{aligned}$$

Additionally, we can also compute $Pr_1^{Ad}[t_1 \cap t_2 \cap E]$ as

$$\begin{aligned} Pr_1^{Ad}[t_1 \cap t_2 \cap E] &= \sum_{I \in LI | I \supset t_1, t_2} Pr_1^{Ad}[I] = (1-\tau) \sum_{I \in LI | I \supset t_1, t_2} Pr_1^{Ad'}[I] \\ &= (1-\tau) Pr_1^{Ad'}[t_1 \cap t_2 \cap E] \end{aligned}$$

As $Pr_1^{Ad'}[E] = 1$, we get $Pr_1^{Ad'}[t_1 \cap t_2] = Pr_1^{Ad'}[t_1] Pr_1^{Ad'}[t_2]$. Also, the prior of $Ad'$ is such that $\sum_{t \in S} Pr_1[t] = 1$. Thus the adversary is indeed $d$-exclusive.